\documentclass[runningheads]{llncs}
\usepackage{graphicx}
\begin{document}
\title{Ontology based Approach \\
        for Precision Agriculture}
%
%
\author{Quoc Hung Ngo \and Nhien-An Le-Khac \and Tahar Kechadi}
%

%
\institute{School of Computer Science, College of Science\\
University College Dublin, Belfield, Dublin 4, Ireland\\
\email{hung.ngo@ucdconnect.ie},\\
\email{an.lekhac@ucd.ie},\\
\email{tahar.kechadi@@ucd.ie}}
\maketitle             
\begin{abstract}
In this paper, we propose a framework of knowledge for an agriculture ontology which can be used for the purpose of smart agriculture systems. 
This ontology not only includes basic concepts in the agricultural domain but also contains geographical, IoT, business subdomains, and other knowledge extracted from various datasets. 
With this ontology, any users can easily understand agricultural data links between them collected from many different data resources. 
In our experiment, we also import country, sub-country and disease entities into this ontology as basic entities for building agricultural linked datasets later.

\end{abstract}

\section{Introduction}

The Internet of Things (IoT) is growing quickly in developing the smart grids, such as
home, health care, transportation, and environment systems as well as smart cities. According to a new forecasts update of International Data 
Corporation (IDC) \cite{IDCltd2018}, worldwide spending on IoT reaches \$772 Billion 
in 2018; an increase of 11.5\% over the \$674 billions that were spent in 2017, 
and this number is predicted to be over \$1 trillion by 2020. Moreover, 
60 percent of global manufacturers will use data analytics from connected 
devices to analyze processes and identify optimization possibilities.

Similarly, IoT in agriculture also grows quickly to improve farm productivity and increase farm profitability. IoT applications in agriculture include vehicle tracking, farm and livestock 
monitoring, storage monitoring, and much more in producing food products \cite{jayaraman2016internet}. 
Considering the future vision of the food lifecycles are well recorded from 
seeds, cultivation, products, transportation, food processing, sales in supermarket, 
it is exciting to have public confidence on food security and high added 
value to the agriculture and food suppliers.

\begin{figure}[htbp]
 \centering
 \includegraphics[width=10.5cm]{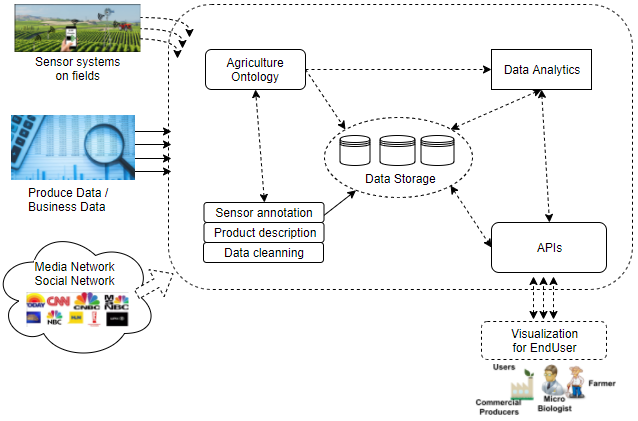}
 \caption{Architecture of Agriculture System}
 \label{figAgriSystem}
\end{figure}

IoT systems are playing an increasingly important role in smart farms, allowing different
organizations and information technology facilities create different datasets. These different datasets create enormous challenges to
integrate them into a workable system so that the midstream firms and the end
consumers can query the history of the agricultural products unhindered by the
bounds of the previous vendors. One of the most important requirements of such
integration is that the data semantics are not consistent among different phases of products. As shown in Figure \ref{figAgriSystem}, to achieve such target, an unified ontology 
of agriculture should be utilized by all the information systems of the 
different phases. Ontology is a concept that is emerging from the various Semantic Web
initiatives, which can be defined as a semantic system that contains terms, 
the definitions of those terms, and the specification of relationships among those terms.

In this research, we propose an agriculture ontology for the purpose of smart
agriculture systems, namely AgriOnt. This ontology will describe basic concepts in the agricultural domain and related thematic subdomains. To keep the ontology light-weight, we ignore the
complexity of the specific agriculture activities or food processing. Only the
environments of the agriculture products are considered so that all the
history can be queried quickly and easily.

The next section gives an overview of an open access linked data, AGROVOC, 
and its related knowledge base in the agriculture domain. Section 3 describes in details our AgriOnt,
Agricultural Ontology. Then, Section 4 discusses the results of several experiments 
on AgriOnt. Finally, we conclude the paper and give some future work in Section 5.

\section{AGROVOC and Related Knowledge Base}

There are several ontologies in precision agriculture. The most popular ontology is AGROVOC. 
AGROVOC\footnote{http://aims.fao.org/standards/agrovoc/linked-data} is a well-known vocabulary system 
that has international interoperability and it consists of over 32,000 concepts available in over 20 
languages \cite{caracciolo2013}. The AGROVOC is aligned from 16 vocabularies related to agriculture and 
is published and managed by the Food and Agriculture Organization of the United Nations (FAO). 
Detailed information about AGROVOC thesaurus is available from the FAO website, while the RDF version 
of AGROVOC can be downloaded at: \\
\textit{http://aims.fao.org/aos/agrovoc/} or \textit{https://datahub.io/dataset/agrovoc-skos} .

However, AGROVOC is only a good vocabulary system to start building an agricultural ontology rather than being used as an ontology for agriculture because some of its relations are inconsistently assigned and others are too broadly defined. For example, there are some insufficient features that can be used as core vocabulary. First of all, the relationship between concepts is not clear or well defined. Most of narrower and broader relationship is attached only by considering the pair-wise relationship. Thus hierarchy by these relationships are not so consistent \cite{joo2016agriculture}. This vague relationship between concepts can lead to the difficulty when adding new terms. In addition, it is difficult to define relations for concepts in AGROVOC, i.e., to find the best position for new terms. Secondly, the number of activity names about rice farming are insufficient. These disadvantages come from the combination of all vocabularies into one vocabulary like AGROVOC.

\begin{figure}[htbp]
 \centering
 \includegraphics[width=12cm]{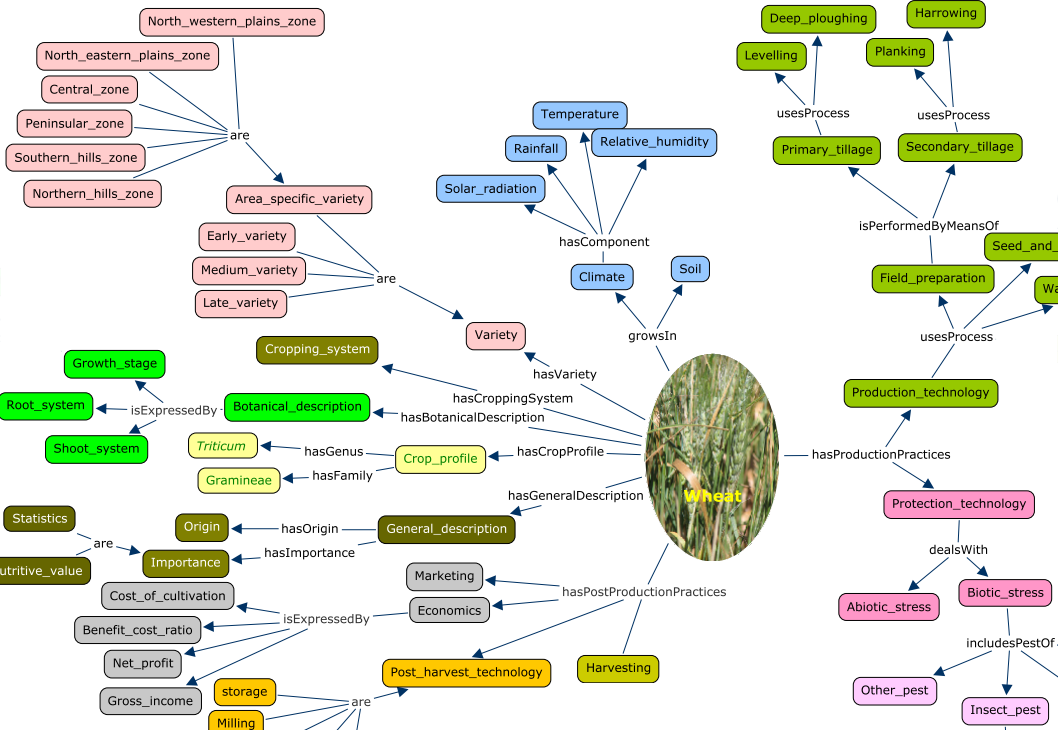}
 \caption{A Screenshot of Knowledge Model for Wheat}
  \label{figureAgropediaWheat}
\end{figure}

Adding to AGROVOC, there are several smaller ontologies for agricultural studies. For example, Plant Ontology of Laure Cooper, et al. \cite{Laure2012}, Animal Ontology, and Animal Disease Ontology of Marie-Colette Fauré, et al. \cite{Faure2011} are built for living entities in agriculture. They are very useful database resources for all plant and animal scientists. According to \cite{Laure2012}, the Plant Ontology consists of over 1,300 rigorously-defined ontology terms and their relations that describe plant anatomy, morphology and developmental stages. Its relations include \textit{partOf} and \textit{hasPart} pair, \textit{precedes} and \textit{precededBy} pair, {participates} and \textit{hasParticipant} pair, in which \textit{dry\_seed\_stage} concept is \textit{precededBy} some seed maturation stages. Moreover, more specific ontologies like Crop Ontology \cite{matteis2013crop} or knowledge models of AgroPedia\footnote{http://agropedia.iitk.ac.in} \cite{sini2009knowledge} (supported and certified by the Indian Council of Agricultural Research) defined very specific concepts and relations (as a screen shot shown in Figure \ref{figureAgropediaWheat} for knowledge model of wheat).

For environmental monitoring, IoT architectures use a collection of numerous active physical things, sensors, actuators, cloud services, and communication layers. These architectures base on SSN ontology to describe sensors and their measurement processes, observations \cite{compton2012ssn}. When applying IoT architectures into agriculture, IoT applications still need a general ontology in the agriculture domain. This ontology can link entities in agriculture (such as animals, plants, fertilizer, diseases), producing aspects (such as weather or soil conditions), and their observations. In fact, HuSiquan, et al. \cite{zhu2015extracting}, or Roussey Catherine, et al. \cite{roussey2010ontologies} also introduced ontologies for agriculture domain. However, these ontologies only contain basic concepts in agriculture, such as \textit{Farm}, \textit{Product}, \textit{Crop}, or \textit{Condition}. Therefore, these studies will be a brief ontology to build a bigger ontology containing all aspects of the agriculture domain.

To summarise, there are several available resources as knowledge-bases in specific topics of the agriculture domain, however, building an agricultural ontology is necessary for applying data science into agricultural management and improving crop yields. Firstly, for example, this knowledge-base can be used to build a data schema for a data warehouse \cite{thenmozhi2014ontological}. Secondly, it can link available resources based on its hierarchy and semantic relations. 

\section{AgriOnt: Agricultural Ontology}

Based on needs of a knowledge-base for precise agricultural applications, an agricultural ontology has an important role in developing its applications. According to Nengfu Xie \cite{xie2007ontology}, an intelligent agricultural knowledge-based service system has an agriculture-specific ontology and a method for agricultural knowledge acquisition and representation. Nengfu Xie and coauthors \cite{xie2007ontology} also describe ontology developing progress which includes three steps: \\
\indent (1) Building a domain-specific knowledge hierarchy; \\
\indent (2) Defining slots of the categories and representing axioms; \\
\indent (3) Knowledge acquisition filling in the value for slots of instances.

\begin{figure}[htbp]
 \centering
 \includegraphics[width=7cm]{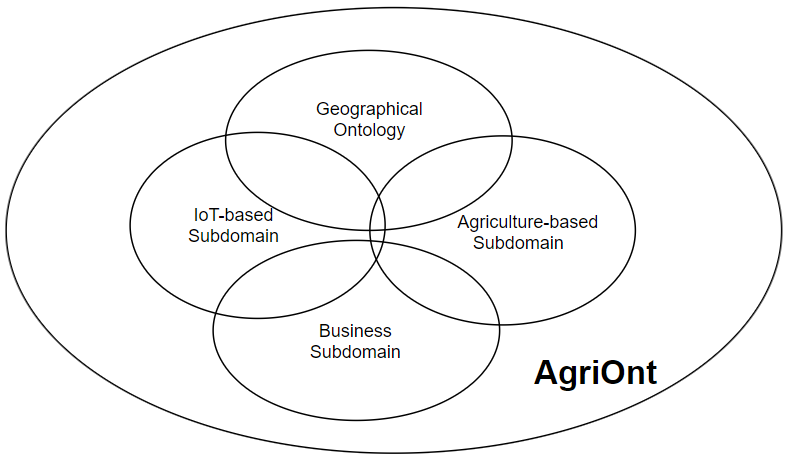}
 \caption{Components of Agricultural Ontology}
  \label{figureAgriOntIdea}
\end{figure}

In general, our new agricultural ontology includes four thematic subdomains: agriculture part, geographical ontology, IoT subdomain, and business subdomain (as shown in Figure \ref{figureAgriOntIdea}). Concepts or classes in each subdomains are inherited from a general class, \textit{Entity}, and two its sub-classes (\textit{VirtualEntity} and \textit{PhysicalEntity}).

\textbf{Entity}: In several ontologies, \textit{Entity} is also called \textit{Thing} and it includes two sub-classes, \textit{VirtualEntity} and \textit{PhysicalEntity} (as shown in Figure \ref{figureAgriOnt}).

\subsection{Agricultural Subdomain}

\begin{figure}[htbp]
 \centering
 \includegraphics[width=12cm]{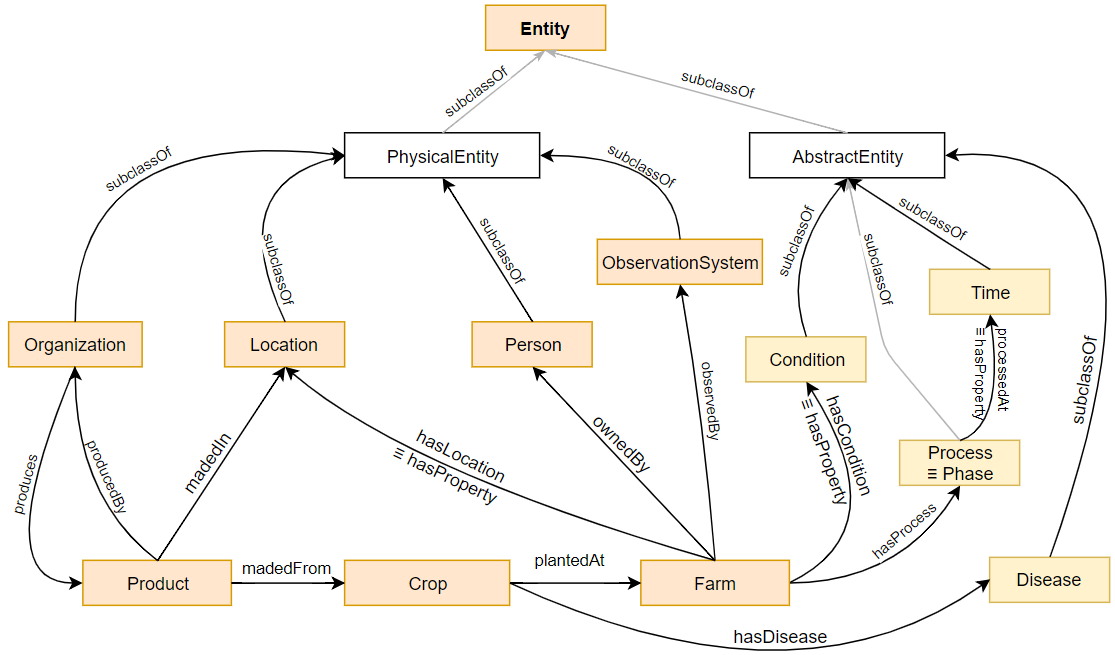}
 \caption{An overview of Agricultural Ontology architecture}
  \label{figureAgriOnt}
\end{figure}

Agricultural subdomain includes basic classes in agriculture domain, such as \textit{Farm}, \textit{Crop}, \textit{Product}, \textit{Fertilizer}, or \textit{Condition}. Figure \ref{figureAgriOnt} shows an overview of agricultural ontology with its main concepts and relationships.

\textbf{Farm} (also called \textit{Field}) mentions a place for planting crops or feeding animals.

\textbf{Product} and its sub-classes are used to describe agricultural products.
Main sub-classes of \textit{Product} are \textit{Food} (\textit{DairyFood} and \textit{ProcessedFood}), \textit{Oil} (\textit{AnimalOil} and \textit{PlantOil}), and \textit{Nutrient}.

\textbf{Crop}, \textbf{Livestock} and their sub-classes are agricultural classes and entities which make products, such as \textit{Cereal}, \textit{Flower}, \textit{Fruit}, \textit{Vegetable} (Crop), \textit{Poultry}, \textit{Cattle} (Livestock), \textit{Fishery}. These concepts can be built based on Plant Ontology\footnote{http://www.plantontology.org}, Animal Ontology\footnote{http://www.cs.man.ac.uk/\~rector/tutorials/Biomedical-Tutorial/Tutorial-Ontologies/Animals/Animals-tutorial-complete.owl}.

\textbf{Process} or \textbf{Phase} class is used to capture positions in the lifecycle of agricultural products. This class has sub-classes, such as \textit{SoilProcess}, \textit{Plainting}, \textit{Spraying}, \textit{Fertilizering}, \textit{Harvesting}, \textit{Marketing}, and \textit{Transportation}.

\begin{figure}[ht]
 \centering
 \includegraphics[width=8cm]{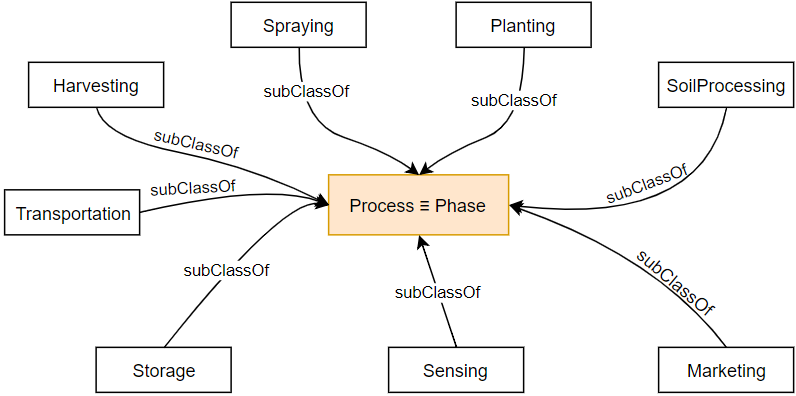}
 \caption{Process and its sub-classes}
 \label{digramPhrase}
\end{figure}

\begin{figure}[ht]
 \centering
 \includegraphics[width=8cm]{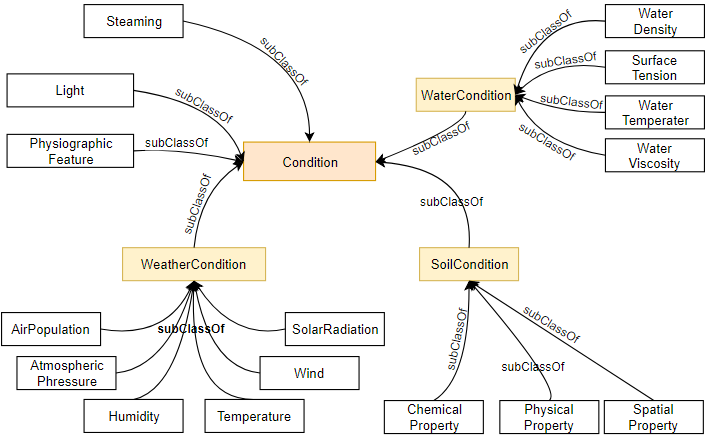}
 \caption{Condition and its sub-classes}
 \label{digramCondition}
\end{figure}

\textbf{Condition} includes everything related to agricultural conditions for producing, such as weather, soil, water, or physiographic features (as shown in Figure \ref{digramCondition}). These conditions are implemented into \textit{WeatherCondition}, \textit{SoilCondition}, \textit{WaterCondition} classes and their features, such as wind speed, temperature, humidity value, chemical properties, and physical properties of soil.

\subsection{IoT Subdomain}
The role of IoT subdomain is connecting sensor systems and linking to observed objects. For this purpose, the Semantic Sensor Network\footnote{https://www.w3.org/2005/Incubator/ssn/ssnx/ssn} (SSN) ontology is a suitable choice to extend and integrate into an agriculture ontology. SSN ontology was developed by W3C Semantic Sensor Networks Incubator Group \cite{compton2012ssn}, and provides a schema that describes sensors, observation, data attributes, and other related concepts at
https://www.w3.org/2005/Incubator/ssn/ssnx/ssn. By using the SSN ontology as a framework to implement IoT aspects into an agricultural system, main classes for this subdomain are:

\textbf{ObserveSystem} or \textbf{System}: A \textit{System} is a unit of IoT infrastructure, it includes a set of sensors or sub-systems. It observes \textit{FeatureOfInterests}, generates observation values for monitoring \textit{Conditions} (such as weather, water, soil) during processing.

\textbf{Sensor}: A sensor is any entity that can follow a sensing method and thus observe several properties of a \textit{FeatureOfInterest}. Sensors may be physical devices, computational methods, a laboratory setup with a person following a method, or any other thing that can follow a Sensing Method to observe a \textit{Property}.

\textbf{FeatureOfInterest}: A \textit{FeatureOfInterest} is a set of \textit{Properties} and it is considered as a \textit{Condition} object in the agricultural part.

\textbf{ObservationValue}: The value of the result of an \textit{Observation}. An \textit{Observation} has a result which is the output of some sensors, the result is an information object that encodes some values for a \textit{FeatureOfInterest}.

\subsection{Geographical Subdomain}

In this ontology, geographical classes includes two main administrative levels (\textbf{Country} and \textbf{Subcountry} classes) and free control locations (\textbf{geo:SpatialThing} and \textbf{geo:Point}). Relationships are longitude/latitude, population, area properties, part-whole relationships among geographical instances, and \textit{hasLocation}, \textit{isLocationOf}, and \textit{isProducedAt} relationships between geographical instances and other instances (plants, animals, products, etc). In fact, a \textbf{Country} is a political division that is identified as a national entity. It also has an unique ISO 3166-1 code in the ISO 3166 standard published by the International Organization for Standardization (ISO). A \textbf{Subcountry} is a subdivision (e.g., province, city or state) of all countries coded in ISO 3166-1, and most of \textit{Subcountry} instances also have unique ISO 3166-2 codes in the ISO 3166 standard. For other geographical instances, they have basic relations, such as longitude, latitude, address, and postcode. They also have \textit{hasCountry} and \textit{hasSubCountry} relationships to mention that they belong with \textit{Country} and \textit{Subcountry} instances.

\subsection{Business Subdomain}

When building agriculture ontology, linking agricultural subdomain and IoT subdomain is necessary to monitor production. This connection can be implemented by \textbf{Organization}, \textbf{Person}, \textbf{Farm} classes and their sub-classes, such as \textit{Company}, \textit{GovernmentOrganization}, \textit{NonGovernmentOrganization} (sub-class of \textit{Organization}), \textit{Farmer}, \textit{LandOwner}  (sub-class of \textit{Person}) classes. Relationships between these classes and \textit{Product} class include \textit{hasProduct}, \textit{produces}, \textit{isProducedBy}, \textit{isProducedAt} relations. With these classes and relationships, the ontology can describe factors contributed to produce agricultural products.

\section{Experiments}

With three main steps in building a core ontology (as mentioned in Section 3), we have built an agriculture ontology with 447 classes and over 700 axioms related to agriculture (as shown in Table \ref{tableOntologyMetrics}). It not only provides an overview of the agriculture domain but also describes agricultural concepts, and lifecycles between seeds, plants, harvesting, transportation, and consumption. It also gives relationships between agricultural concepts and related concepts, such as weather, soil conditions, fertilizers, farm descriptions.

\begin{table}[htbp]
\centering
  \caption{Ontology metrics}
  \label{tableOntologyMetrics}%
    \begin{tabular} {lrrr}
    \hline\noalign{\smallskip}
         	                &          & \textbf{with}     & \textbf{with} \\
        \textbf{Figure} & \hspace{1.5cm}\textbf{Core} & \hspace{0.5cm}\textbf{Geo-data} & \hspace{0.5cm}\textbf{Diseases} \\
    \hline\noalign{\smallskip}
        \textbf{Axiom}  	                & 1843	& 64,805    & 108,062  \\
        \textbf{Logical axiom count} 	    &  749  & 50,876	& 75,316   \\
        \textbf{Declaration axioms}         & 728   & 9,240     & 15,951   \\
        \textbf{Class count} 	            & 447   & 447	    & 447     \\
        \textbf{Object property count}      & 69    & 69	    & 69     \\
        \textbf{Data property count}        & 27    & 27        & 27	 \\
        \textbf{Individual count}           &  101  & 8,615	    & 15,392 \\
    \hline\noalign{\smallskip}
    \end{tabular}%
\end{table}%

To provide basic geo-location data in this ontology, we extract countries and sub-countries, and then import them into our AgriOnt ontology based on studies of Quoc Hung, et al. \cite{Nigel2007}\cite{quochung2011}\cite{quochung2012}\cite{sondoan2008Geospatial}. Most of them contain ISO 3166-1 codes for country level instances and ISO 3166-2 codes for sub-country level instances. Geographical data (as shown in Table \ref{tableGeographial}) also contains longitude, latitude, population, area, agricultural land area, climate condition information, and Wikipedia links.

\begin{table}[htbp]
\centering
  \caption{Detail of Geographial part}
  \label{tableGeographial}%
    \begin{tabular}{llr}
    \hline\noalign{\smallskip}
        \textbf{Entry} 	& \textbf{Detail}& \textbf{Count} \\
    \hline\noalign{\smallskip}
        \textbf{Country}  	& With ISO 3166-1 code			& 243 \\
        \textbf{Sub-country}& With ISO 3166-2 code			& 4,085	 \\
         				    & Without ISO 3166-2 code 		& 142	 \\
        \textbf{Relations} 	& Longitude, Latitude, ISO code, Wikipedia,  	& 23,991	 \\
        				    & population, area, climate				& \\
    \hline\noalign{\smallskip}
    \end{tabular}%
\end{table}%

For smart agriculture systems, chemical and biological control of plant and animal diseases is remarkably high. Therefore, agricultural ontologies contain knowledge-bases about diseases and related sectors is necessary. In this ontology, animal disease instances are imported from Animal Disease Ontology\footnote{http://lovinra.inra.fr/2015/09/28/maladies-animales/} of Fauré Marie-Colette and Aubin Sophie \cite{Faure2011} while plant disease instances are extracted from Plant Disease pages of APS Journals\footnote{https://apsjournals.apsnet.org}.

With existing resources, we have built an agricultural ontology with geographical instances (countries and sub-countries), diseases, micro-organisms. Moreover, we also manually collect and create main instances in the agricultural domain, such as crops, animals, and related typical products.

\begin{figure}[ht]
 \centering
 \includegraphics[width=12cm]{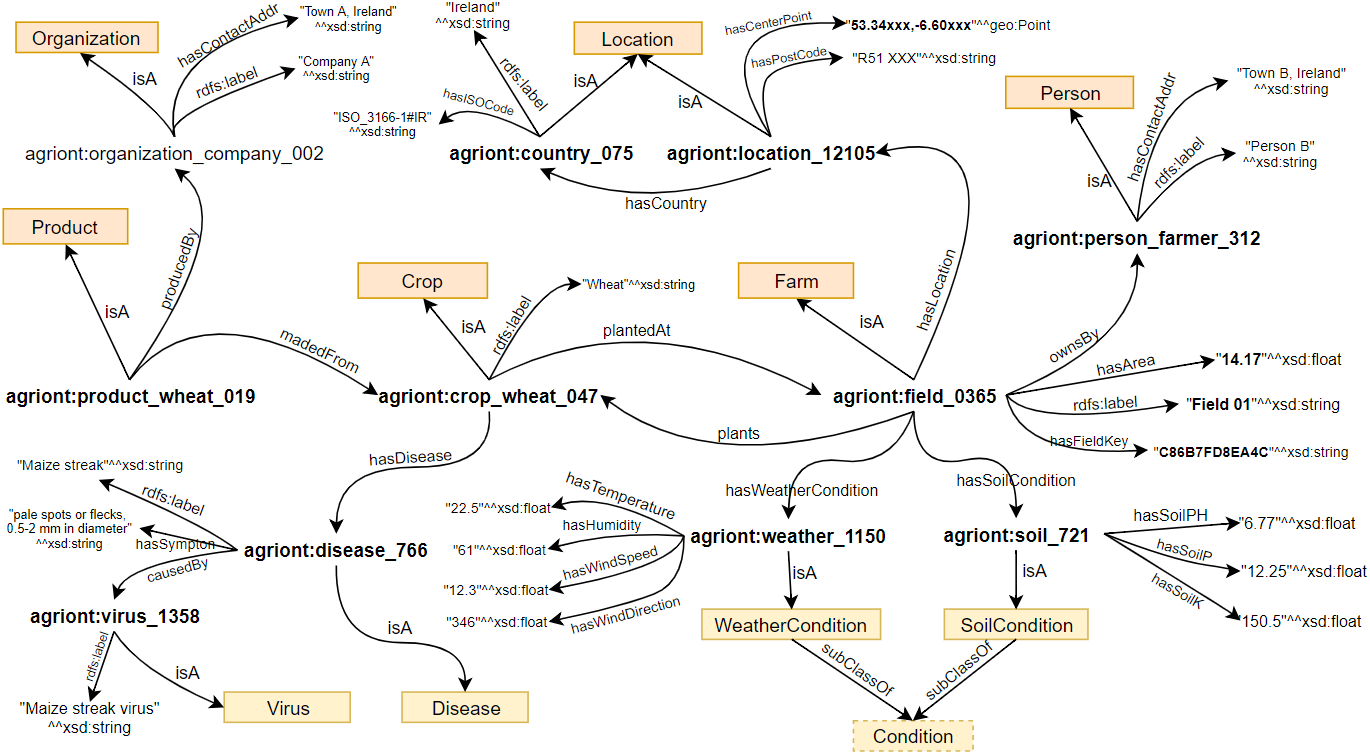}
 \caption{Example of linked data based on AgriOnt.}
 \label{AgriOntSample}
\end{figure}

In our scenario, this ontology will be a semantic framework to build a linked data for agriculture applications and analytics. As shown in Figure \ref{AgriOntSample}, the process relates to products, crops, farms, farmers, diseases, affected aspects and producing conditions. In fact, this linked data example shows features of field \textit{"Field 01"} (has URI \textit{agriont:field\_0365}), such as its basic characteristics, locaton, weather condition (\textit{agriont:weather\_1150}), and soil condition (\textit{agriont:soil\_721}).

Moreover, ontologies can be used to build data warehouse schema. According to M. Thenmozhi \cite{thenmozhi2014ontological}, database tables can be inherited ontology classes while attributes of tables are mapped to properties of the equivalent classes or relationships between classes. In our scenario, \textit{Product}, \textit{Crop}, \textit{Farm} or \textit{Field}, \textit{Farmer}, \textit{SoilCondition}, and \textit{Weather} are considered to be database tables in the data warehouse schema and their descriptions and relations will help to define features of database tables \cite{vuong2018warehouse}. Similarly,  relationships between concept classes are presented as relations between database tables in whole schema of the data warehouse.

\section{Conclusion and Future Work}
We presented an ontology architecture for agriculture domain. We describe four main thematic subdomains of an agricultural ontology and give descriptions of typical concepts of each subdomain. 
The architecture of our proposed ontology with these four sub-domains aims to tackle the challenges in pre-processing and analysing real-world argriculture datasets. In addition, raw argriculture data collected from big enterprises today is quite fit in Big Data context with its volume, variety and velocity. Normally, the preprocessing of these datasets requires data reducing techniques \cite{lekhac2010datareduction} that may cause the missing of important information. Using this proposed ontology can assist the process of integrating, harmonising and transformation of raw agriculture data efficiently.      
Furthermore, this ontology can be applied to real-world datasets to store as linked data and apply data analytics techniques. Current studies on agricultural ontology mostly focus on vocabulary and concept models of this domain, but our research shows that agricultural ontologies can be used to manage produce progress and analyse agricultural data as well. For example, researchers can use AgriOnt to integrate weather data into agricultural datasets as the weather condition which is one of the factors that affects crop yields. Moreover, this ontology with agricultural hierarchy can help to integrate available resources to
build larger and precise knowledge maps \cite{lekhac2007knowledgemap}.

\section*{Acknowledgement}
This research forms part of CONSUS and is funded under the SFI Strategic Partnerships Programme  (16/SPP/3296)
and is co-funded by Origin Enterprises Plc


\begin{thebibliography}{1}

\bibitem{caracciolo2013}
Caracciolo, Caterina, Armando Stellato, Ahsan Morshed, Gudrun Johannsen, Sachit Rajbhandari, Yves Jaques, and Johannes Keizer. 
"The AGROVOC Linked Dataset". 
Semantic Web 4, no. 3 (2013), pp. 341-348.

\bibitem{compton2012ssn}
Compton Michael, Barnaghi Payam, Bermudez Luis, Garc{\'\i}A-Castro Ra{\'u}L, Corcho Oscar, Cox Simon, Graybeal John, Hauswirth Manfred, Henson Cory, Herzog Arthur, and others. 
"The SSN ontology of the W3C semantic sensor network incubator group". 
Web semantics: science, services and agents on the World Wide Web, 2012, 17: 25-32.

\bibitem{Faure2011}
Faure M. C., Zundel E., Aubin S., and Millox M. (2011). 
"The animal diseases reference system: from CERISA to Linked Data".
In: Animal Health: News and Quality (pp.97). 
Presented at Scientific Animation Days of the Department of Animal Health, Fréjus, France, 22-25, May, 2011 (in French).

\bibitem{IDCltd2018}
International Data Corporation (IDC) Ltd. 
"Forecasts Worldwide Spending on the Internet of Things". 
URL: https://www.idc.com/getdoc.jsp?containerId=prUS43295217, publish on 07 Dec 2017, access on 30 April 2018.

\bibitem{jayaraman2016internet}
Jayaraman Prem Prakash, Yavari Ali, Georgakopoulos Dimitrios, Morshed Ahsan, and Zaslavsky Arkady. 
"Internet of things platform for smart farming: Experiences and lessons learnt".
Sensors 16.11 (2016): 1884.

\bibitem{joo2016agriculture}
Joo Sungmin, Koide Seiji, Takeda Hideaki, Horyu Daisuke, Takezaki Akane, and Yoshida Tomokazu. 
"Agriculture Activity Ontology: An Ontology for Core Vocabulary of Agriculture Activity".
International Semantic Web Conference (Posters \& Demos). 2016.

\bibitem{kawtrakul2012ontology}
Kawtrakul, Asanee.
"Ontology engineering and knowledge services for agriculture domain". 
Journal of Integrative Agriculture 11.5 (2012): 741-751.

\bibitem{heshan2016}
Heshan Du, Vania Dimitrova, Derek Magee, Ross Stirling, Giulio Curioni, Helen Reeves, Barry Clarke, and Anthony Cohn. 
"An ontology of soil properties and processes".
International Semantic Web Conference. Springer, Cham, 2016.

\bibitem{zhu2015extracting}
Hu Siquan, Wang Haiou, She Chundong, and Wang Junfeng.
"AgOnt: ontology for agriculture internet of things".
International Conference on Computer and Computing Technologies in Agriculture. Springer, Berlin, Heidelberg, 2010.

\bibitem{Laure2012}
Laure Cooper, Ramona L. Walls, Justin Elser, Maria A. Gandolfo, Dennis W. Stevenson, Barry Smith, Justin Preece, Balaji Athreya, and others. 
"The plant ontology as a tool for comparative plant anatomy and genomic analyses". 
Plant Cell Physiol. 54(2): e1, 2012, pp. 1–23. doi:10.1093/pcp/pcs163.

\bibitem{matteis2013crop}
Luca Matteis, Pierre-Yves Chibon, Herlin Espinosa, Milko Skofic, Richard
Finkers, Richard Bruskiewich, Glenn Hyman, and Elizabeth Arnaud. 
"Crop ontology: vocabulary for crop-related concepts", 
The First International Workshop on Semantics for Biodiversity (S4BioDiv), 2013.

\bibitem{sini2009knowledge}
Margherita Sini, Vimlesh Yadav, Jeetendra Singh, Prabhakar TV, and Vikas Awasthi.
"Knowledge models in agropedia indica". 
FAO, 2009

\bibitem{lekhac2010datareduction}
N-A. Le-Khac, M. Bue, M. Whelan and M-T.Kechadi
"A clustering-based data reduction for very large spatio-temporal datasets". 
International Conference on Advanced Data Mining and Applications,(ADMA’2010), Chongquing, China, November 2010, Springer Verlag LNAI Vol.6441: 43-54 

\bibitem{lekhac2007knowledgemap}
N-A.Le-Khac, L. Aouad and M-T. Kechadi 
"Distributed knowledge map for mining data on grid platforms". 
International Journal of Computer Science and Network Security (2007), Vol.7(10):98-107

\bibitem{xie2015research}
Nengfu Xie, Wensheng Wang, Bingxian Ma, Xuefu Zhang, Wei Sun and Fenglei Guo. 
"Research on an agricultural knowledge fusion method for big data". 
Data Science Journal, 2015, 14, pp.2-9.

\bibitem{xie2007ontology}
Nengfu Xie, Wensheng Wang, and Yong Yang. 
"Ontology-based agricultural knowledge acquisition and application".
Proceedings of the International Conference on Computer and Computing Technologies in Agriculture. Springer, Boston, MA, 2007,  pp. 349-357.

\bibitem{Nigel2007}
Nigel Collier, Ai Kawazoe, Son Doan, Mika Shigematsu, Kiyosu Taniguchi, Lihua Jin, John McCrae, Hutchatai Chanlekha, Dinh Dien, Quoc Hung, Van Chi Nam, Koichi Takeuchi, DEng, Asanee Kawtrakul.
"Detecting Web Rumours with a Multilingual Ontology-Supported Text Classification System", 
Journal Advances in Disease Surveillance, Volume 4, pp. 242.

\bibitem{quochung2011}
Quoc Hung Ngo, Son Doan, and Werner Winiwarter. 
"Building a Geographical Ontology by Using Wikipedia". 
In Proceedings of the 13th International Conference on Information Integration and Web-based Applications \& Services, pp. 345-348. ACM, 2011

\bibitem{quochung2012}
Quoc-Hung Ngo, Son Doan, and Werner Winiwarter. 
"Using Wikipedia for extracting hierarchy and building geo-ontology". 
International Journal of Web Information Systems 8.4 (2012): 401-412.

\bibitem{roussey2010ontologies}
Roussey Catherine, Soulignac Vincent, Champomier Jean-Claude, Abt Vincent, and Chanet Jean-Pierre. 
"Ontologies in agriculture".
AgEng 2010, International Conference on Agricultural Engineering. Cemagref, 2010, pp. 1-10.

\bibitem{sondoan2008Geospatial}
Son Doan, Quoc Hung Ngo, Nigel Collier.
"Building and Using Geospatial Ontology in the BioCaster Surveillance System", 
Workshop on Bio-Ontologies 2008: Knowledge in Biology, July 2008.

\bibitem{thenmozhi2014ontological}
Thenmozhi, M., and K. Vivekanandan. 
"An ontological approach to handle multidimensional schema evolution for data warehouse." 
International Journal of Database Management Systems 6.3 (2014): 33.

\bibitem{vuong2018warehouse}
Vuong M. Ngo, Nhien-An Le-Khac, and Tahar M. Kechadi. 
"An Efficient Data Warehouse for Crop Yield Prediction." 
14th International Conference on Precision Agriculture, International Society of Precision Agriculture, 2018.

\bibitem{zheng2012construction}
ZHENG, Ye-lu, et al.
"Construction of the ontology-based agricultural knowledge management system". 
Journal of Integrative Agriculture 11.5 (2012): 700-709.



\end{thebibliography}
\end{document}